\documentclass[paper,nofootinbib]{revtex4}

\usepackage{graphicx}
\usepackage{epsfig}
\usepackage{fancyhdr}
\usepackage{empheq}
\usepackage{mathbbol}
\usepackage{wasysym}
\usepackage{bbm}
\usepackage{bm}
\usepackage{color}

\usepackage[utf8x]{inputenc}
\usepackage{amssymb,amsmath}
\usepackage{epstopdf}
\usepackage{float}
\usepackage{color}
\usepackage{subfig}

\newcommand{\be}{\begin{equation}}
\newcommand{\ee}{\end{equation}}
\newcommand{\bea}{\begin{eqnarray}}
\newcommand{\eea}{\end{eqnarray}}

\begin{document}

\title{Bounds to the Higgs Sector Masses in Minimal Supersymmetry from LHC Data}
\author{L Maiani$^{1,3}$, AD Polosa$^{1}$} 
\author{V Riquer$^2$}
\affiliation{$^1$Dipartimento di Fisica, Sapienza Universit\`a di Roma, Piazzale A Moro 5, Roma, I-00185, Italy\\
$^2$Fondazione TERA, Via G Puccini 11, I-28100 Novara, Italy\\
$^3$ CERN, PH-TH, 1211 Gen\`eve 23, Switzerland}

\begin{abstract}
We update our analysis of the  Higgs sector masses in Minimal Supersymmetry and in the Two-Higgs Doublet  Model  to the final ATLAS and CMS data from the 2011-2012 LHC run.
\newline\newline
PACS: 12.60.Jv, 14.80.Cp, 14.80.Ec
\end{abstract}

\maketitle

\thispagestyle{fancy}

\paragraph*{{\bf Introduction.}} 
The search for positive indications of low energy Supersymmetry is one of the challenges of present LHC experiments. Realistic  models with broken Supersymmetry, however, admit many free parameters so as to make the search for the supersymmetric companions of the known particles a rather complex affair.

One univocal prediction of Supersymmetry is the need for an extended Higgs field structure, with respect to the Standard Theory: at least two $SU(2)\otimes U(1)$ doublets in its simplest version, the Minimal Supersymmetric Standard Model (MSSM). 

In a previous paper~\cite{Maiani:2012_I} we stressed  that, with the  newly discovered boson $h(125)$~\cite{ATLAS,CMS} identified with the lightest MSSM Higgs boson, the physics of the Higgs sector has a very mild dependence upon the parameters of the full supersymmetric theory. The leading radiative corrections may be fixed by the $h$ mass so that the Higgs sector physics is described by two parameters only, the ratio of the two vacuum expectation values, conventionally denoted as $\tan\beta$, and the mass of one of the two other neutral Higgs bosons, $H$~($J^P=0^+$) or $A$~($J^P=0^-$), for a recent discussion see~\cite{djrecent}. 

In a second paper~\cite{Maiani:2012_II}, a comparison was made of the Higgs sector MSSM description with the first data on the production and decay modes of  $h(125)$ made available by the ATLAS and CMS collaborations at CERN~\cite{ATLAS,CMS} and by the CDF and D0 collaborations at the Tevatron~\cite{tevatron}. A consistent fit was obtained and lower bounds to the mass of $H$ were derived, of $320$ ($2\sigma$)  and $360$~GeV ($1\sigma$), with the masses of $A$ and $H^\pm$ essentially degenerate with $H$ in this mass range. The theoretical decay branching fractions of $H$ and $A$ were reported as functions of $M_H$ and $\tan\beta$. A picture of partial decoupling of the additional Higgs sector with $\tan\beta=1- 10$ seems viable. The partial decoupling scenario goes well with the picture of a relatively light scalar top envisaged in~\cite{Delgado:2012eu} and is not inconsistent with the lack of observed deviations, so far, in flavour changing neutral currents from the predictions of the Standard Theory (ST), see {\it e.g.}~\cite{isidoristraub,nazila}. 

In the present paper we update our analysis to the final ATLAS and CMS data from the 2011-2012 LHC run~\cite{ATLAS2,CMS2}, also extending the MSSM analysis to the more general Two-Higgs Doublet Model (THDM). The extended Higgs sector of the Next to Minimal Supersymmetric Model (NMSSM) has been compared to present data in~\cite{Barbieri:2013hxa,belanger}.

\vskip0.2cm

\paragraph*{{\bf Couplings.}}
We denote, as usual, by $H^0_d$ and $H^0_u$ the $0^+$ neutral scalar fields and define $0\leq \beta \leq \pi/2$ according to:
\bea
&&\langle0|H^0_u|0\rangle=v \sin\beta \;\;\;\;\;  \langle0|H^0_d|0\rangle=v \cos\beta \;\;\;\;\;  0< \tan\beta< +\infty     \label{vevs} \\
&& v^2=(2 \sqrt{2} G_F)^{-1}=(174\;{\rm GeV})^2
\label{standardvev}
\eea 
 For conserved $CP$ symmetry, the physical $0^+$ particles are orthogonal combinations of these two fields, determined by a second angle $\alpha$. The two angles determine the couplings of the Higgs scalars to quarks, leptons and gauge bosons. In formulae:
 \bea
&&h= S_{hi} H_i \;\;\;\;\;\;\; H=S_{Hi}H_i \;\;\;\;\; ( i=d,u)\notag \\ 
&& S_{hd}=\cos\alpha\;\;\;\;\; S_{hu}=\sin\alpha
\label{mixalfa}
\eea
The ratios of the couplings of $h$ and $H$ to the ST couplings for the different channels are summarised in Tab.~\ref{couplings}.

 \begin{table}[hb]%
\label{tab:couplings}\centering%
\begin{tabular}{|lc|l|clc|l|c}
\hline%
 && $WW=ZZ$&& $t\bar t = c\bar c$ && $b\bar b =\tau^{+} \tau^{-}$\\
\hline
$h$ && $ \cos(\beta-\alpha) $ && $\sin\alpha( \sin\beta)^{-1}$ && $\cos\alpha (\cos\beta)^{-1}$ \\
\hline
$H$ && $ \sin(\beta-\alpha) $ && $-\cos\alpha( \sin\beta)^{-1}$ && $\sin\alpha (\cos\beta)^{-1}$ \\
\hline
\end{tabular}
\caption{
}
\label{couplings} 
\end{table}
 
We may require the coupling to vector bosons to be positive and $\alpha$ is restricted to the range:
\be
\beta-\pi/2<\alpha<\beta+\pi/2
\label{alfarange}
\ee

 The line $\alpha=\beta$ in the ($\beta$, $\alpha$) plane maps into the ST point $c_t=c_b=c_V=1$ in coupling constant space.
  
  
  The above equations suggest to fit the cross section and decay rates data of $h(125)$ with three independent couplings~\cite{Maiani:2012_II}, $c_t,c_b,c_V$ for the $up$ quarks and leptons, $down$ quarks and leptons and for the vector boson, respectively, rather than with a single normalization factor (denoted by $\mu$ in Refs.~\cite{ATLAS,CMS}) or with one fermion and one vector coupling~\cite{ATLAS2,CMS2}\cite{contino,Ellis:2013lra}. 
 \vskip0.2cm
\paragraph*{{\bf The MSSM mass matrix.}}

The structure of the scalar mass matrix in MSSM has been studied extensively (see {\it e.g.} Ref.~\cite{susybasic}). We start from the expression: 
 \bea
&&{\cal M}_S^2=(M_Z^2+\delta_1)
\left(\begin{array}{cc}  \cos^2\beta & -\cos\beta\sin\beta \\ -\cos\beta\sin\beta & \sin^2\beta \end{array}\right) + M_A^2\left(\begin{array}{cc}  \sin^2\beta & -\cos\beta\sin\beta \\ -\cos\beta\sin\beta & \cos^2\beta \end{array}\right) 
+\left(\begin{array}{cc} 0 & 0 \\0 &\frac{\delta}{\sin^2\beta} \end{array}\right) 
\label{massmatrix}
\eea

Radiative corrections due to $b$ and other lighter quarks are neglected, which is justified for the relatively small values of $\tan\beta$ we shall consider. The corrections embodied by $\delta_1$ and $\delta$ arise from $t$ quark and $t$ scalar quarks loop and  are given approximately by~\cite{Carena:1995wu}:
\bea
\delta_1&=&-\frac{3}{8 \pi^2}~\frac{M_t^2M_Z^2}{v^2}~t ;~\label{delta1corr}
\\
\delta&=&\frac{3}{4 \pi^2}~\frac{M_t^4}{v^2}\left[\frac{1}{2} {\tilde X}_t + t +\frac{1}{16 \pi^2} (\frac{3 M_t^2}{2 v^2} - 32 \pi \alpha_S)( {\tilde X}_t  +t)t\right]~\label{deltacorr} \\
t&=&\log\left(\frac{M_{\text{SUSY}}^2}{M_t^2}\right)
\label{corrctns}
\eea 
where $M_{\text{SUSY}}^2=M_{{\tilde t_1}} M_{{\tilde t_2}} $ fixes the scale of supersymmetry breaking and ${\tilde X}_t$ is a mixing parameter~\cite{Carena:1995wu} with:
\be
{\tilde X}_t<6
\label{range}
\ee
   

We note that $\delta_1$ is a subdominant correction. In fact, $\delta_1={\cal O}(g^2y_t^2)$ while $\delta={\cal O}(y_t^4)$, with $g$ and $y_t$ the electroweak and the $t$-quark Yukawa couplings, see Eqs.~(\ref{delta1corr}) and (\ref{deltacorr}). Numerically, we find $|\delta_1/\delta| \leq 0.25$ in the range $0.5\leq M_{\text{SUSY}}\leq 10$~TeV and $0\leq {\tilde X}_t\leq 6$.  Within this precision\footnote{
alternatively, one may assume a given value of ${\tilde X}_t$, express both $\delta_1$ and $\delta$ as functions of $t$, Eq. (\ref{corrctns}), and use the secular equation of $M_h$ to eliminate $t$. In this case all results depend upon the assumed value of ${\tilde X}_t$. In the range (\ref{range}), we find the results to be insensitive to the value of 
${\tilde X}_t$ and numerically indistinguishable from the results obtained by neglecting $\delta_1$.}, we shall neglect $\delta_1$. 

One obtains $\delta$ as function of $M_A$ and $\tan\beta$ from the secular equation for $M_h$, without any assumption on the mixing parameter ${\tilde X}_t$. 

One finds, explicitly:

\be
\frac{\delta}{\sin^2\beta}= \frac{(M_A^2-M_h^2)(M_h^2-M_Z^2)+M_A^2 M_Z^2 \sin^2(2\beta)}{\sin^2\beta M_A^2-(M_h^2-M_Z^2+\sin^2 \beta M_Z^2)}
\label{deltavalue}
\ee
with the large $M_A$ limit:
\be
M_h^2=M_Z^2\cos^2(2\beta) +\delta+{\cal O}\left(\frac{M_Z^2}{M_A^2}, \frac{M_h^2}{M_A^2}\right)
\label{fixeigen1}
\ee

Eq. (\ref{deltavalue}) has a singularity for:
\be
M_A^2=\frac{M_h^2-M_Z^2}{\sin^2 \beta}+ M_Z^2
\label{singul}
\ee
which implies a lower bound for $M_A^2$, since, below this value, the second eigenvalue, $M_H^2$, becomes negative. Within the validity of Eq. (\ref{massmatrix}) with $\delta_1$ neglected, we find, numerically:
\bea
&& M_A\geq 155~{\rm GeV},~ (\tan\beta=1),\notag \\
 && M_A\geq M_h,~ (\tan\beta=10)
\label{lowbnd}
\eea

Substituting Eq. (\ref{deltavalue}) into (\ref{massmatrix}), we obtain $M_H$ and the mixing coefficients (\ref{mixalfa}) as functions of $M_A$ and $\tan\beta$.

 The couplings of $h$ to the $up$ and  $down$ fermions and to the vectors are given by: 
 
 \bea
c_t(\tan\beta,m_H)&=&\frac{\sqrt{1+\tan^2\beta}}{\tan\beta}\;S_{hu}(\tan\beta,m_H)\\
c_b(\tan\beta,m_H)&=&\sqrt{1+\tan^2\beta}\; S_{hd}(\tan\beta,m_H)\\
c_V(\tan\beta,m_H)&=&\frac{1}{\sqrt{1+\tan^2\beta}}\; S_{hd}(\tan\beta,m_H)+\frac{\tan\beta}{\sqrt{1+\tan^2\beta}}\; S_{hu}(\tan\beta,m_H)
 \eea
\vskip0.2cm

\paragraph*{{\bf Three-couplings fit to the $h(125)$ data.}}

It is straightforward to express the ST decay widths of $h$ in terms of the parameters ($c_t$, $c_b$, $c_V$):
\be
\Gamma(h\to b \bar b)=c_b^2~ \Gamma(h\to b \bar b)_{\text{ST}},~\dots
\ee
we refer to~\cite{Maiani:2012_I,Maiani:2012_II} for details. We compute the ST widths with the code HDECAY~\cite{Hdecay}. 

The $t$ quark couplings appears in the $h\to\gamma\gamma$ loop amplitude,  together with $c_V$, so that the decay width is sensitive to their relative sign. 
In MSSM there are also contributions from the scalars ${\tilde t_1} $ and ${\tilde t_2}$ which renormalize the $t$ quark coupling: $c_t\to\tilde c_t$, making it different from the coupling of the $up$-like quarks. For a mass of the lighter scalar $t$ quark around $0.5$~TeV, the correction amounts~\cite{Maiani:2012_II} to about $3\%$ and, in any case, the contribution from $u$ and $c$ quarks to the decay width are entirely negligible, so we may for the moment disregard this effect.
        
The total width, necessary to obtain the branching fractions,  is computed by adding individual widths in the ST channels.  
There may be invisible channels in $h$ decay, {\it e.g.} due to light SUSY neutralinos~\cite{ATLAS:2013pma}. This would appear as a reduction of the couplings obtained from the fit  by a common factor $f<1$.

The $h$ production cross section receives two contributions. A dominant one due to gluon-gluon fusion, which scales with $c_t$ and the subdominant Vector Boson  Fusion, which scales with $c_V$\footnote{interference among g-g and VBF amplitudes is discussed extensively in the literature and known to be negligible.
}.  We have computed the ST cross sections for $p+p\to H + 2~jets$
using Alpgen~\cite{alpgen} and applied the experimental cuts on jet energy and on the diphoton rapidity, for the $p+p\to h\to \gamma \gamma$ case, spelled out by the CMS collaboration~\cite{CMS}. Similar cuts apply to the ATLAS case
,  see ~\cite{Maiani:2012_II} for details.

With the new data in~\cite{ATLAS2,CMS2} and Tevatron data~\cite{tevatron} we find one absolute minimum 
with positive $c_t$:
\be
 {\bar c}_t = 0.99 \;\;\;\;\;{\bar c}_b = 1.23\;\;\;\;\;{\bar c}_V = 1.11\;\;\;\;\; (\chi^2=11.7,~\chi^2/{\rm dof}=1.46)
 \label{positive}
\ee
and a slightly unfavored local minimum:
\be
{\bar c}^{(-)}_t = -1.19\;\;\;\;{\bar c}^{(-)}_b = 0.57\;\;\;\;{\bar c}^{(-)}_V = 0.50\;\;\;\;(\chi^2=12.1,~\chi^2/{\rm dof}=1.51)
\ee
\begin{figure}[htb]
\begin{minipage}[t]{80mm}
\includegraphics[scale=.6]{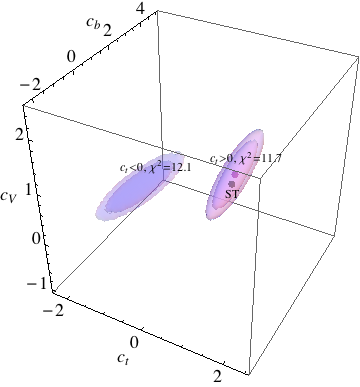}       
\caption{\label{fig:soluz1} {\small Positions of the two minima of the three-dimensional fit, the $1\sigma$ and $2\sigma$ regions, and the Standard Theory point. The solution with $c_t<0$ is only a local minimum of $\chi^2$.}}
\end{minipage}
\hspace{\fill}
\begin{minipage}[t]{75mm}
\includegraphics[scale=.5]{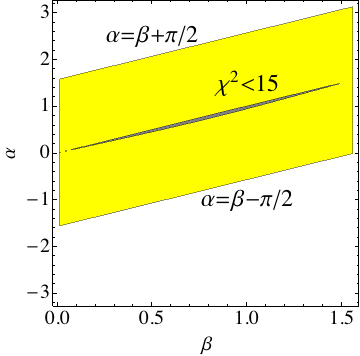}   
\caption{\label{fig:fit2param} {\small  The $\chi^2$ fit to the two parameters, $\alpha,\beta$ of the Two Higgs Doublets Model (THDM). The dark shaded region represents the distribution of the points with  $\chi^2<15$.}}
\end{minipage}
\end{figure}
We illustrate in Fig.~\ref{fig:soluz1}  Positions of the two minima of the three-dimensional fit, the $1\sigma$ and $2\sigma$ regions, and the ST point, which lies well inside the $1\sigma$ region of the absolute minimum, corresponding to  $c_t>0$. 

For comparison with other work, we have made a fit with $c_t=c_b=c_F$. In this case, the difference of $\chi^2$ between the absolute minimum ($c_t>0$,  $\chi^2=12.6$) and the local minimum ($c _t<0$, $\chi^2=14.6$) is more pronounced.

We have performed also a $\chi^2$ fit to the two parameters, $\alpha,\beta$ of the Two Higgs Doublets Model (THDM). The distribution of the points with  $\chi^2<15$ is reported in Fig.~\ref{fig:fit2param}. Points cluster along the ST line $\alpha=\beta$, confirming the results in \cite{Grinstein:2013npa}, see also~\cite{Eberhardt:2013uba,Giardino:2013bma}.

\begin{figure}[htb]
\begin{minipage}[t]{80mm}
\includegraphics[scale=.55]{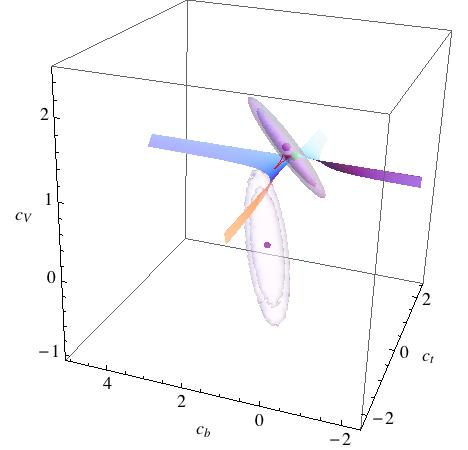}       
\caption{\label{fig:soluz2}{\small The THDM $(\beta,\alpha)$ surfaces embedded in the  coupling constants space. The small triangle hitting the ST point represents the restriction of the THDM to Supersymmetry. }}
\end{minipage}
\hspace{\fill}
\begin{minipage}[t]{75mm}
\includegraphics[scale=.55]{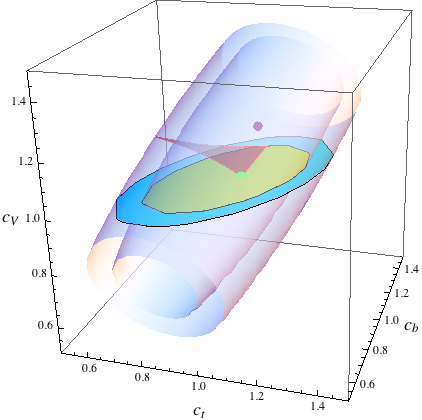}   
\caption{\label{fig:velain3d} {\small The small triangle in Fig.~\ref{fig:soluz2} is essentially flat on the plane $c_V=1$ and lies below the positive solution in~(\ref{positive}).}}
\end{minipage}
\end{figure}

The situation in coupling constants space is illustrated in Fig.~\ref{fig:soluz2}. The two surfaces hitting the positive solution in the ST point represent the ($\beta$, $\alpha$) region, with the range of $\alpha$ in (\ref{alfarange}) reduced to $\alpha-\beta=\pm\pi/6$, in view of the result in Fig.~\ref{fig:fit2param}. The small triangle hitting the ST point represents the restriction of the THDM to Minimal Supersymmetry. The triangle is essentially flat, on the plane $c_V=1$, below the positive solution in (\ref{positive}), as shown in Fig.~\ref{fig:velain3d}.

\begin{figure}[htb]
\begin{center}
\includegraphics[scale=.65]{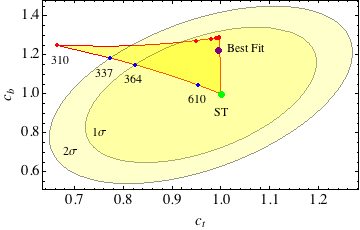}       
\caption{\label{fig:ellissi}  {\small The projection of the allowed SUSY region on the $c_V=1$ section with the $1\sigma-2\sigma$ regions.  The sides of the triangle pointing to the Standard Theory point correspond to $\tan\beta=1$, on the left, and $\tan\beta\geq10$, on the right and $310 \leq M_H \leq 800$~GeV. Marked on the line $M_H=310$~GeV are points with $\tan\beta=1,2.5,4,\cdots 10$. }}
%
%
\end{center}
\end{figure}

Our main results are reported in Fig.~\ref{fig:ellissi}, where we give the projection of the allowed SUSY region on the $c_V=1$ section, with the $1\sigma-2\sigma$ regions. The ST limit is indicated, as well as the projection of the best fit solution. The sides of the triangle pointing to the ST point correspond to $\tan\beta=1$, on the left, $\tan\beta\geq10$, on the right and $ M_H= 310$~GeV. Marked on the line $M_H=310$~GeV are points with $\tan\beta=1,2.5,4,\cdots 10$. On the left side of the triangle, we have indicated the $1\sigma- 2\sigma$ points. We conclude that:
\bea
&&\tan\beta=1\notag \\
&&M_H> 340~{\rm GeV},~(2 \sigma)\notag \\
&&M_H> 360~{\rm GeV},~(1 \sigma)
\label{limit1}
\eea
and:
\bea
&&\tan\beta\geq10\notag \\
&&M_H> 330~{\rm GeV},~(1 \sigma)
\label{limit2}
\eea
These values are considerably larger than the bounds in Eq.~(\ref{lowbnd}) and, more important, they are consistent with and slightly larger than the experimental lower bounds derived by ATLAS and CMS~\cite{Aad:2012yfa,CMS:gya}. Unfortunately, there are no upper bounds to $M_H$ since, for $M_H$ increasing, the SUSY point is driven towards the ST point, which is well within $1\sigma$ from the best fit.

\paragraph*{\bf Conclusions.}
The intermediate decoupling region, with $\tan\beta=1- 10$, $M_H$ in the few hundreds GeV region and  a scalar top below $1$~GeV, is not excluded by the data on $h(125)$ and by Flavor Changing Neutral Currents data.
The three couplings fit to the data is very effective for mapping the allowed region of the heavier scalars predicted by MSSM. The lower bounds: $M_H\leq340 (360)$~GeV are derived at $2\sigma$ ($1\sigma$) confidence level, with $A$ and $H^\pm$ essentially degenerate. These values are consistent with recent experimental bounds by ATLAS and CMS~\cite{Aad:2012yfa,CMS:gya}. The fit is sensitive to the experimental cuts. A fit produced by the LHC collaborations with the precise experimental cuts they have adopted would be extremely useful.

\end{document}